\documentclass[prl,twocolumn,showpacs,superscriptaddress,nofootinbib,preprintnumbers]{revtex4-2}
\usepackage[bottom]{footmisc} 

\usepackage{graphicx}
\usepackage{subcaption}
\usepackage{amsmath}
\usepackage{mathrsfs}
\usepackage{xcolor}
\usepackage{hyperref}
\usepackage{cleveref}
\usepackage{comment}
\usepackage{gensymb}
\usepackage{style}

\crefname{equation}{Eq.}{Eqs.}
\crefname{figure}{Fig.}{Figs.}

\usepackage{dcolumn}
\usepackage{bm}

\hypersetup{colorlinks=true, linkcolor=blue, citecolor=blue, urlcolor=blue, filecolor=blue}

\newcounter{qnumber}
\newcommand{\fy}[1]{{\color{blue} [FY:\,#1]}}

\newcommand{\vecp}{\bold{p}}

\newcommand{\vecq}{\bold{q}}

\begin{document}

\title{Heating Up the Black Hole X-ray Binary Accretion Disk by Superradiance}

\author{Antonios Kyriazis\,\orcidlink{0000-0002-7351-1691}} 
\email{akyriazis@ufl.edu}
\affiliation{Department of Physics, University of Florida, Gainesville, FL 32611,  U.S.A.}
\affiliation{Department of Physics and Astronomy, University of Iowa, Iowa City, IA, 52242, USA}

\author{Fengwei Yang\,\orcidlink{0000-0001-9873-6259}}
\email{fyang7@nd.edu}
\affiliation{Department of Physics and Astronomy, University of Notre Dame, South Bend, IN 46556}

\author{Siyu Zhou\,\orcidlink{0009-0000-2758-4251}}
\email{szhou8@nd.edu}
\affiliation{Department of Physics and Astronomy, University of Notre Dame, South Bend, IN 46556}

\begin{abstract}
   A superradiant cloud of ultralight axions around a black hole, that is part of an X-ray binary system, can heat up its accretion disk and be detected by the thermal X-ray spectrum emitted by the disk. We consider a derivative coupling of the axions to the plasma fermions and calculate the emissivity of the inverse bremsstrahlung process that results in a temperature fluctuation of the disk. Based on the thin-disk model and the multicolor disk model, we derive the thermal spectrum with axion heating, which shows an enhanced thermal photon flux and a red-/blue- shifted peak spectral frequency. 
   A single bump hunting search of the axion heating signature in the thermal spectrum of a $10M_\odot$ black hole X-ray binary with a spectral measurement sensitivity of 10\% (1\%) can derive the constraint on axion-electron coupling $|g_{ae}|\gtrsim7.5\times 10^{-12} ~(2.4 \times 10^{-12})$ for axion mass $m_a=5.2\times 10^{-12}\,$eV in a saturated $|211\rangle$ state, and  $|g_{ae}|\gtrsim4.5\times 10^{-12} ~(1.4 \times 10^{-12})$ for axion mass $m_a=1.0\times10^{-11}\,$eV in a saturated $|322\rangle$ state. The projected sensitivities are competitive with those from XENONnT. 
   A detailed continuum fitting can further improve the detectability and provide a complementary bound to the black hole spin-down measurement. 
   
\end{abstract}

\maketitle

\section{Introduction}
Black hole (BH) superradiance (also known as the Penrose process) is a process that can amplify incident waves on rotating BHs\,\cite{1969NCimR...1..252P,1972JETP...35.1085Z,Bekenstein:1973mi,Misner:1972kx,Teukolsky:1974yv,Bekenstein:1998nt}.
For the ultralight axion or axion-like particle fields\,\cite{Arvanitaki:2009fg,Hui:2016ltb}, they can be gravitationally bound to the BH in large numbers, forming a superradiant cloud.
If axion-to-Standard-Model couplings are assumed, such as axion-photon coupling, interesting phenomena can follow \cite{Spieksma:2023vwl,Blas:2020nbs,Rosa:2017ury,Boskovic:2018lkj,Sen:2018cjt}. 
One of the dominant processes is parametric resonance production of photons 
that leads to large amounts of energy released by the cloud. 
The converted photons obtain plasma mass in the interstellar medium around the BH, making the conversion process 
available only if the axion mass is greater than the plasma mass. 
The emitted photon frequency from a stellar-mass BH is $\sim\,$kHz
(see though \cite{Blas:2020nbs} for an indirect method using $\textrm{CMB}$ observations).      

\begin{figure*}[h!t]
        \centering
            \includegraphics[width=0.45\linewidth]{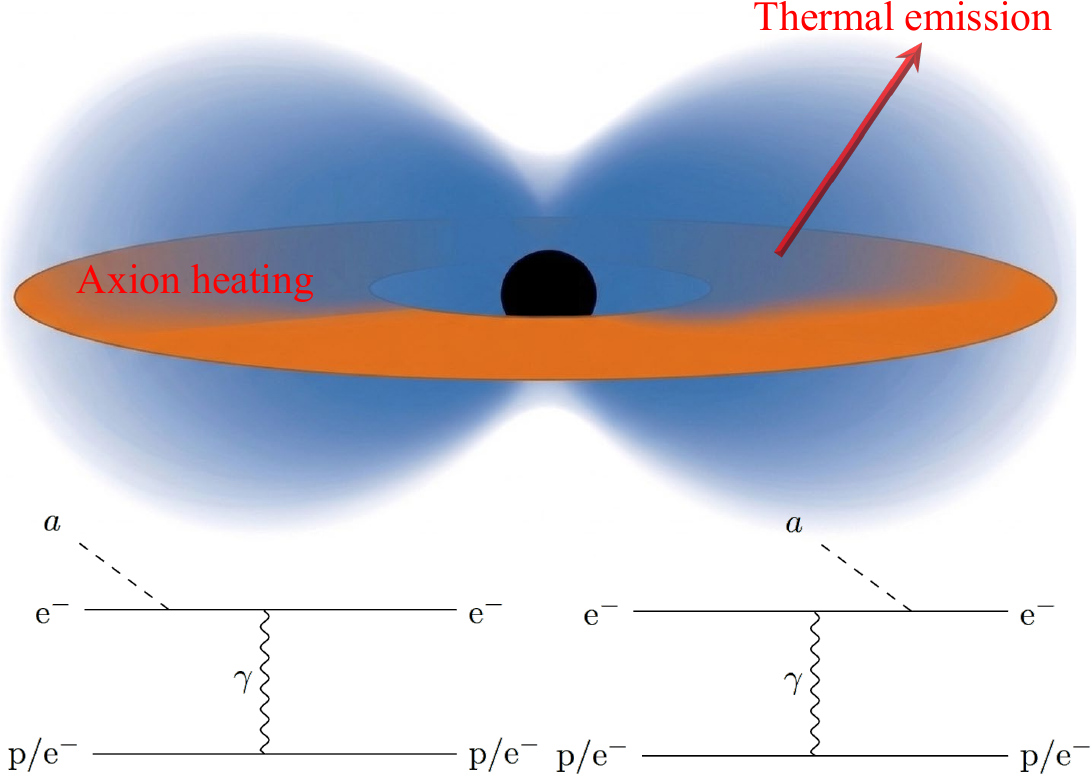}\quad\includegraphics[width=0.45\linewidth]{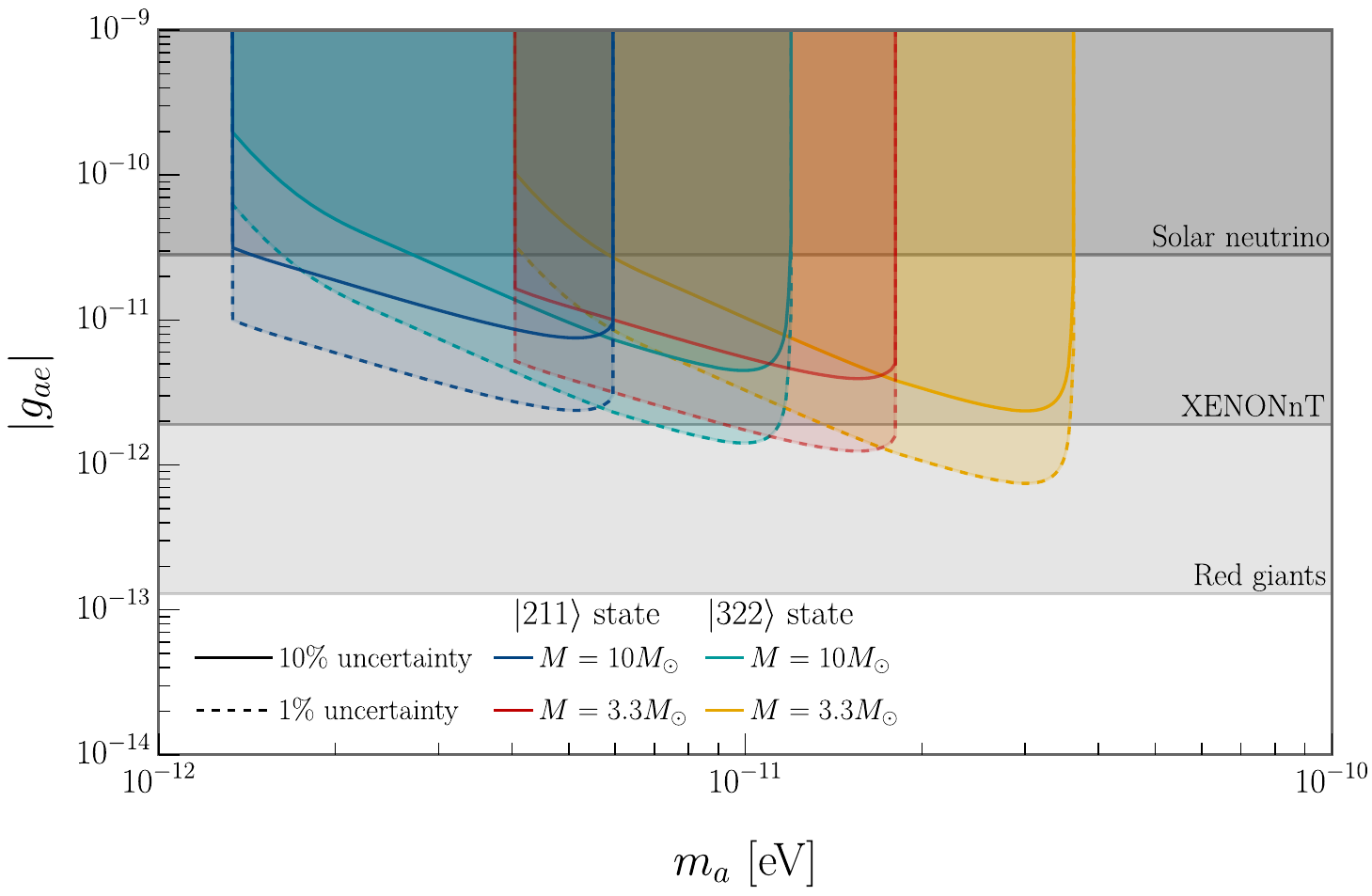}
    \caption{ {\bf Left top}: The schematic plot of the system with a BH X-ray binary accretion disk and an axion cloud in $|211\rangle$ state. {\bf Left bottom}: The relevant Feynman diagrams of inverse bremsstrahlung via axion-electron coupling.  
    {\bf Right}: The upper limit of axion-electron coupling $|g_{ae}|$ as a function of axion mass $m_a$ shown in blue and red shaded regions. We use the SuperRad to obtain axion cloud profile for $|211\rangle$ state ($\alpha\in[0.1,\,0.445]$) and $|322\rangle$ state ($\alpha\in[0.1,\,0.896]$), assuming BH initial spin $\tilde{a}_{\rm ini}=0.995$, and final spin $\tilde{a}_{\rm f}$ is defined by saturated spin. The current constraints from solar neutrinos \cite{Gondolo:2008dd}, XENONnT \cite{XENON:2022ltv}, and Red giants \cite{Capozzi:2020cbu} are shown in the gray shaded region.  }
    \label{fig:MCD_spec_bm1}
\end{figure*}

Axion-fermion coupling, widely studied in the context of stellar cooling 
\cite{1986PhRvD..33..897R,Raffelt:1996wa,Carenza:2021osu,Ning:2025tit,Buschmann:2019pfp,Buschmann:2019pfp,Isern:2008fs,Isern:2008nt,Viaux:2013lha,MillerBertolami:2014rka,Krauss:1984gm}, however, lacks exploration in the presence of an axion cloud. In this case, the inverse process is relevant: axions from the superradiant cloud are absorbed by the ambient electrons, increasing their kinetic energy and heating them up. We apply this idea to the accretion disk of an X-ray binary (XRB) system, a region with a high concentration of electrons, making it an ideal environment to study interactions with axions and the resulting observational signatures.

An XRB  consists of a stellar-mass BH accreting material from its companion. The accretion flow can form an optically thick and geometrically thin disk. The accretion efficiently converts the gravitational energy into radiation, emitting strong electromagnetic radiation, predominantly in the X-ray band. The radiation from the disk is locally that of a blackbody, but with a temperature that decreases from the inner to the outer region, and thus the thermal spectrum from the entire disk is described by a multicolor blackbody spectrum \cite{Shakura:1972te,1973blho.conf..343N,Thorne:1974ve,Page:1974he}. Fitting the X-ray observation data with a multicolor disk (MCD) model provides precise BH spin measurement \cite{2006ApJ...652..518M,2009ApJ...701L..83S,2021cosp...43E1412R}. 

Accretion can affect the superradiance mechanism in a number of different ways. Firstly, if the cloud is unable to develop due to the BH mass being too small, accretion can increase its mass and facilitate the superradiance process.
Secondly, after the cloud has grown, the BH's spin remains at its saturated value,
which is a time-dependent quantity, given that the BH's mass increases. The system is then found in an equilibrium, as any increase of the spin is compensated by superradiance, while any decrease is compensated by accretion. This defines a particular trajectory in the Regge plane, spanned by the BH mass and its spin, and has been used to derive constraints on axions from highly spinning BHs\,\cite{Brito:2014wla,Guo:2022mpr,Hui:2022sri,Nandakumar:2025aex,Guo:2025dkx,Li:2026gup}. Given these well-established results, in what follows, for a given axion and BH mass, we use the corresponding saturated spin. 

In this {\it Letter}, we will derive the emissivity for the inverse-bremsstrahlung process relevant for the axion heating and the temperature fluctuation on the accretion disk,  present the thermal X-ray spectrum including the heating effect, and obtain the projected sensitivities for axion-electron coupling and axion mass. We use natural units $c=\hbar=1$.

\section{Axion heating}
\label{sec: axion heating}

The axion-electron coupling that we are interested in is given by 
\begin{align}
\mathcal{L}_{\rm int} = \frac{{g}_{ae}}{2 m_{e}} \partial_{\mu} a \bar{\psi}_{e} \gamma^{\mu} \gamma^{5} \psi_{e}\,,
\end{align}
where $m_{e}$ is the electron mass, $a$ is the axion field, $\psi_{e}$ is the electron field, and ${g}_{ae} $ is the axion-electron coupling\,\footnote{Another interaction can be considered for the axion-proton coupling. However, since the proton mass is much larger than $m_{e}$, the interaction rate will be suppressed and we shall ignore them.}. For the above interaction, the relevant processes are the inverse bremsstrahlung ($a + e + e/p \rightarrow e + e/p$)  and Compton scattering ($ a + e \rightarrow e + \gamma $). For the former process, the electron interacts, via the exchange of a virtual photon, either with other electrons or with protons or ions in the accretion disk. Compton scattering, 
however, is kinematically forbidden, because the effective mass of final-state photon, given by the plasma frequency $ \omega_{\rm eff} = \sqrt{e^{2} n_{\rm e}/m_{e}} \simeq 3.7\, \textrm{eV}$
for a typical electron density $n_{\rm e}=10^{22} \textrm{cm}^{-3}$ in the disk~\cite{Braaten:1993jw}, is much larger than the axion mass\,\footnote{For the same reason, the process $a \rightarrow \gamma \gamma$, which is relevant when an axion-photon coupling is considered, is forbidden in the accretion disk and that is why we have ignored it.},  
$m_{a} \sim 10^{-12} - 10^{-11} \textrm{eV}$ for a stellar-mass BH. Hence, only inverse-bremsstrahlung is relevant for the axion heating processes.

The squared amplitude of the inverse bremsstrahlung process $a + e + p \rightarrow e + p$ is, 
\begin{align}
\label{eqn:amplitude}
    \frac{1}{4}\sum_{\rm spins}\lvert\mathcal{M}_{\rm ep}\rvert^2 
    \simeq \frac{4 e^{4} {g}^{2}_{ae} m^{2}_{p}|\textbf{q}|^{2}}{m^{2}_{e} \left( |\textbf{q}|^{2} + \omega_{\rm eff} ^{2} \right)^2},
\end{align}
where all particles are non-relativistic,
$\textbf{q} = \textbf{p}_{e,i} - \textbf{p}_{e,f} + \textbf{k}_{a}
$ is the momentum transfer, with $\textbf{p}_{e,i(f)}$ the momentum of the initial (final) electron, $\textbf{k}_{a}$ the momentum of the axion\,\footnote{In Ref.\,\cite{Raffelt:1996wa}, the Debye-Huckel screening scale $k_S$ is also used in the plasmon propagator, but we show that the emissivity is insensitive to the different choices of the IR cutoff in the Coulomb scattering, in which the form factor changes by $\sim\log(m_{\rm e}T/\Lambda_{\rm IR}^2),~\Lambda_{\rm IR}=\{\omega_{\rm eff},k_S\}$, so we just show the result using plasma frequency in the main text. }.

The {\it emissivity} of axion heating following Ref.\,\cite{1986PhRvD..33..897R},
\begin{align}
\epsilon_a&\simeq\frac{1}{T}\int {\rm d}\Pi_aE_a^2{\rm d}\Pi_1{\rm d}\Pi_2{\rm d}\Pi_3{\rm d}\Pi_4 f_af_1f_2(1-f_3)(1-f_4)  \nonumber \\
&\times(2\pi)^4\delta^{(4)}(p_1+p_2+k-p_3-p_4)|\mathcal{M}|^2 \label{eqn: emissivity definition}
\end{align}
where 
${\rm d}\Pi_i\equiv(2\pi)^{-3}(2E_i)^{-1}{\rm d}^3p_i$,  
$f_{1(2)}$,\,$f_{3(4)}$,\,and $f_{a}$ are the phase space distribution functions of the initial, final fermion, and axion, respectively.
By evaluating \cref{eqn:amplitude,eqn: emissivity definition}\,\footnote{See Supplemental Material for detailed derivation.}, we obtain the axion heating rate for electron-proton scattering,
\begin{align}
    \frac{\epsilon_a}{n_{\rm p}}&=\frac{1}{8\pi^3}e^4\left(\frac{g_{a\rm e}}{2m_e}\right)^2\rho_an_{\rm e}\left(\frac{2\pi}{m_{\rm e}T}\right)^{\frac{3}{2}}F(\zeta) \label{eq: energy per electron}\\
    =&\frac{1}{2}\frac{\rm eV}{\rm s}\left(\frac{g_{a\rm e}}{10^{-11}}\right)^2\frac{E_an_a}{10^{41}{\rm eV}{\rm cm}^{-3}}\frac{n_{\rm e}}{10^{22}{\rm cm}^{-3}}\left(\frac{10^7 {\rm K}}{T}\right)^{\frac{3}{2}}F(\zeta),  \nonumber 
\end{align}
where $E_a$ is the axion energy, $n_a$ is the axion number density, $\zeta\equiv\omega_{\rm eff}^2/(8Tm_{\rm e})$, $F(\zeta)=-\left(1+\gamma+\ln{\zeta}\right)/2$ is a form factor that describes the IR divergence of the Coulomb scattering and $\gamma$ is the Euler constant. The heating rate of electron-electron scattering is roughly $1/\sqrt{2}$ of Eq.~(\ref{eq: energy per electron}), 
(see Supplemental Material for more details).
The heating process is proportional to $n_a$ and $n_{\rm e}$ as expected, given that it is a $3$-body scattering process; notably, it is more efficient for lower temperatures\,\footnote{Here, we treat $n_{\rm e}$ as independent of $T$, but in a realistic setup, the overall temperature dependence of the axion heating rate can also depend on $n_{\rm e}(T)$. }.

In this work, we consider a superradiant cloud around the XRB with the gravitational fine-structure constant $\alpha \equiv G M m_{a}=\mathcal{O}(0.1-1)$. For small $\alpha$, the axion number density profile $n_a({\bf r})$ can be described by Hydrogenic states $|n\ell m\rangle$ and $E_a\simeq m_a$ in the non-relativistic limit \cite{Arvanitaki:2010sy,Arvanitaki:2014wva,Baumann:2019eav,Kyriazis:2025fis}. For $|211\rangle$ state, 
in spherical coordinates with a BH at the origin and its spin aligned with the $z$-axis, the wavefunctions we will consider are
\begin{align}
\label{eq:wavefunctions NR}
    \psi_{211}(r,\theta=\frac{\pi}{2},\varphi) & = \frac{1}{8 \sqrt{\pi} r^{3/2}_{c}} \frac{\tilde{r}}{\tilde{r}_{c}} e^{-\tilde{r}/2 \tilde{r}_{c}} e^{i \varphi},   \\     \label{eq:wavefunctions NR 322}
    \psi_{322} (r,\theta = \frac{\pi}{2},\varphi) & = \frac{1}{162 \sqrt{ \pi} r^{3/2}_{c}} \frac{\tilde{r}^{2}}{\tilde{r}^{2}_{c}} e^{-\tilde{r}/3 \tilde{r}_{c}} e^{2 i \varphi},
\end{align}
where $\tilde{r} \equiv r / r_{g}$, with $r_{g} \equiv GM$ the gravitational radius of BH, $\tilde{r}_c= \alpha^{-2}$ the dimensionless Bohr radius of the axion cloud.  The profiles are evaluated at the equatorial plane $\theta=\pi/2$, since we consider a thin disk model. Thus, for a given state, $n_{a} = N_{a} |\psi_{nlm}|^{2} $, where $N_{a}=q_c M/m_a$ is the number of axions in a superradiant state, and $q_{c}$ is the ratio of the cloud's mass to the BH mass $M$. We use the \texttt{SuperRad} package to numerically determine $q_{c}$ for given values of $\alpha$ when superradiance has saturated and the cloud has reached its maximum mass \cite{Siemonsen:2022yyf,May:2024npn}.

The wavefunctions of \cref{eq:wavefunctions NR,eq:wavefunctions NR 322} are invalid for $\alpha\sim 1$, and numerical general-relativistic calculations are needed to obtain accurate results. In what follows, we made use of the \texttt{bhsr} code in \cite{hoof_git}, and expanded it based on the results of \cite{Dolan:2007mj}, to calculate the wavefunctions in the parameter space where the non-relativistic approximation does not hold (see Appendix D of \cite{Kyriazis:2025fis} for comparison between relativistic and non-relativistic wavefunctions).

\section{Accretion Disk Model}
\label{sec:accretion disk model}
In this work, we use the Novikov-Thorne (NT) disk model\,\cite{1973blho.conf..343N}. 
We focus on the {\it inner region} where radiation pressure dominates over gas pressure, and the {\it middle region} where gas pressure is dominant. The opacity of both regions is predominantly due to electron scattering. 
The disk model can be parametrized by two disk model parameters $\{\alpha_d,\,\dot{m}\}$ and two BH parameters $\{\tilde{m},\,\tilde{a}\}$, 
where $\alpha_d$ is the Shakura-Sunyaev viscosity parameter, $\dot{m}\equiv \dot{M}c^2/ L_{\rm Edd}$ 
is the dimensionless accretion rate, 
with the Eddington luminosity $L_{\rm Edd}\equiv 4\pi GMc/ \kappa_{\rm es}$, and the electron scattering opacity $\kappa_{\rm es}\simeq 0.4 \,{\rm cm}^2/{\rm g}$, $\tilde{m}\equiv M/M_\odot$, and $\tilde{a}\equiv a/M$ is the dimensionless BH spin.
Given the input of these parameters, the plasma density profile $ \rho_{\rm pl}$, core temperature profile $T_{\rm c}$, and surface density profile $\Sigma$ can be derived from the disk model, shown as follows \cite{Abramowicz:2011xu}\,\footnote{The profiles quoted here include the latest corrections on the NT disk model.}.
In the {\it inner region}: 
\begin{align}
   \rho_{\rm pl} &\simeq 2\times 10^{-5}\text{g/cm}^3\alpha_d^{-1}\tilde{m}^{-1}\dot{m}^{-2} \tilde{r}^{3/2} h_1(x,\tilde{a})\, ,\label{eq:inner_rho}\\
   T_{\rm c}&\simeq5\times 10^7\,{\rm K}\,\alpha_d^{-1/4}\tilde{m}^{-1/4}\tilde{r}^{-3/8}h_2(x,\tilde{a}),\label{eq:inner_Tc}\\
   \Sigma&\simeq5\text{g/cm}^2\,\alpha_d^{-1}\dot{m}^{-1}\tilde{r}^{3/2}h_3(x,\tilde{a}),\label{eq:inner_Sigma}
\end{align}
and in the {\it middle region}:
\begin{align}
   \rho_{\rm pl} &\simeq 40\,\text{g/cm}^3\alpha_d^{-\frac{7}{10}}\tilde{m}^{-\frac{7}{10}}\dot{m}^{2/5} \tilde{r}^{-\frac{33}{20}}h_4(x,\tilde{a}) ,\label{eq:middle_rho}\\
   T_{\rm c}&\simeq7\times 10^8\,{\rm K}\,\alpha_d^{-\frac{1}{5}}\tilde{m}^{-\frac{1}{5}}\dot{m}^{2/5}\tilde{r}^{-\frac{9}{10}}h_5(x,\tilde{a}),\label{eq:middle_Tc}\\
   \Sigma&\simeq9\times10^4\text{g/cm}^2\,\alpha_d^{-4/5}\tilde{m}^{1/5}\dot{m}^{3/5}\tilde{r}^{-3/5}h_6(x,\tilde{a}),\label{eq:middle_Sigma}
\end{align}
where 
auxiliary functions $h(x,\tilde{a})$ with $x\equiv \sqrt{\tilde{r}}$ are introduced in Supplemental Material, encapsulating relativistic correction coefficients $\mathcal{A,B,C,D,E}$ which asymptote to unity far from the BH, and the relativistic Page-Thorne flux factor $\mathcal{Q}$ that encodes the boundary condition for the disk profile \cite{Page:1974he}. For the parameter space of interest, the radius relevant for axion heating is within the middle region as it will be shown that the axion heating effect is maximal at $\tilde{r}\simeq n^2 \tilde{r}_c$.

The local radiation from the disk arises from the viscosity dissipation \cite{Pringle:1981ds}, and the axion heating at the perturbative level (as shown in Supplemental Material the local thermal equilibrium is always established),    
\begin{equation}
\label{eqn:flux}
    \frac{4}{3\tau}\sigma_{\rm SB} (T_{\rm c}+\delta T_{\rm c})^4=\underbrace{\frac{3 GM\dot{M}}{8\pi r^3}\mathcal{Q}'}_{\rm viscosity ~dissipation}+\underbrace{\frac{1}{2}\frac{\epsilon_a (T_{\rm c})\Sigma(r)}{\rho_{\rm pl}(r)}}_{\rm axion~heating}\,,
\end{equation}
where $\mathcal{Q}'\equiv\mathcal{Q}\mathcal{B}^{-1}\mathcal{C}^{-1/2}$ (inner region) or $\mathcal{Q}\mathcal{B}^{-4/5}\mathcal{C}^{-1/2}$ (middle region) is the flux factor with vertical structure correction \cite{1995ApJ...450..508R}, 
$\tau=\Sigma \kappa$ is the optical depth of the disk, $\kappa=\kappa_{\rm es}$ in both inner and middle region, and $\sigma_{\rm SB}$ is the Stefan-Boltzmann constant, $\delta T_{\rm c}$ is the core temperature fluctuation due to axion heating such that $\delta T_c=0$ if axion heating is absent. The axion heating term denotes the energy injection per unit area per unit time via inverse bremsstrahlung where the $1/2$ prefactor is from the assumption that the fluxes radiated from the both sides of the disk are symmetric.

The temperature fluctuation is obtained by Taylor expanding \cref{eqn:flux} to the leading order,
\begin{align}
\label{eqn:temp fluct}
    \frac{\delta T_{\rm c}}{T_{\rm c}}=\frac{3\tau\epsilon_a \Sigma}{{32} \sigma_{\rm SB} T^4_c\rho_{\rm pl}}\propto q_cg_{ae}^2\alpha^{\frac{9}{5}}M^{-\frac{6}{5}}\dot{m}^{-\frac{3}{5}}\alpha_d^{-\frac{6}{5}}\,,
\end{align}
while the scaling behavior applies to the middle region with $\alpha\lesssim0.3$.

\begin{figure}
    \centering
    \includegraphics[width=0.95\linewidth]{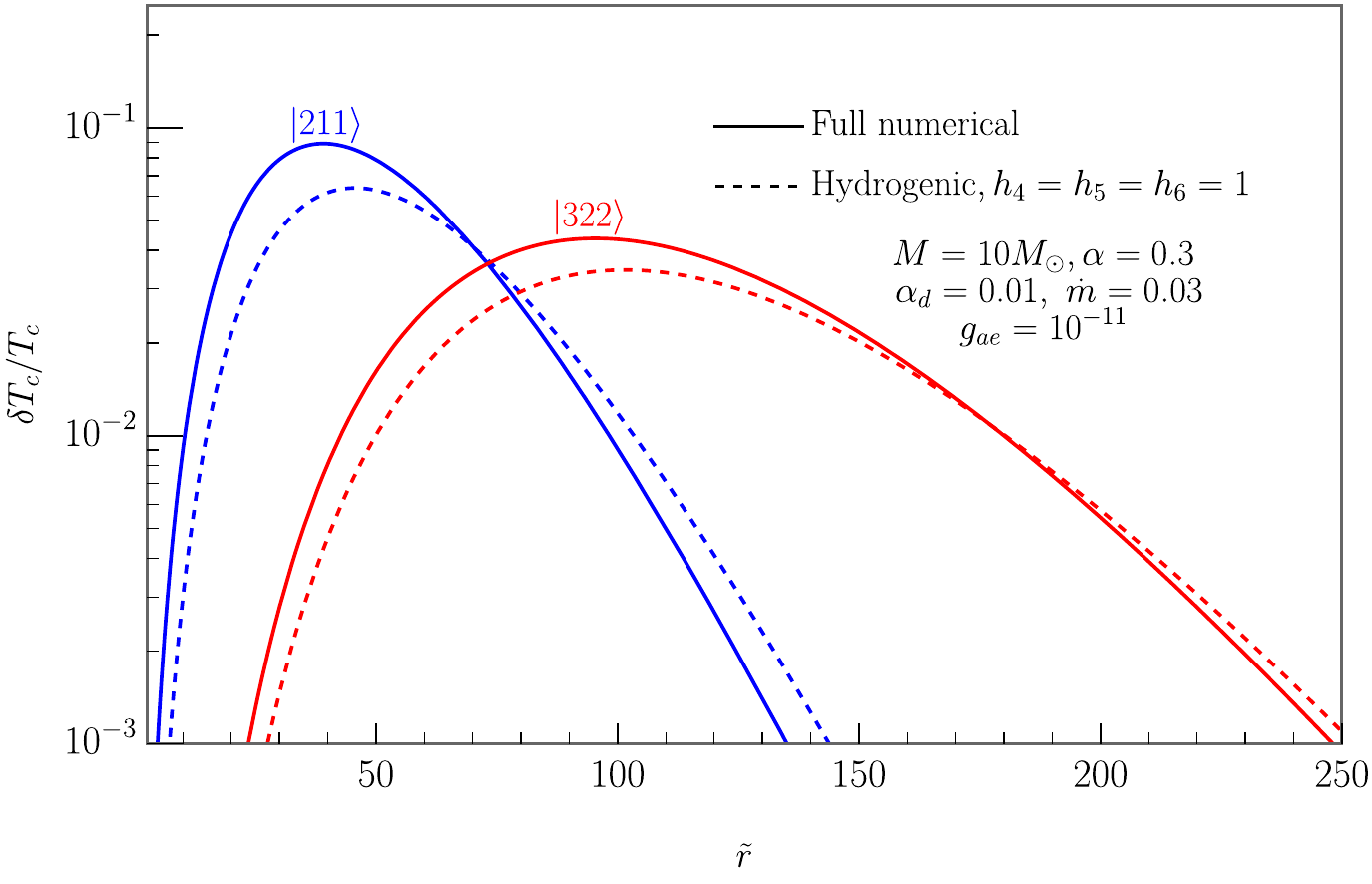}
    \caption{The temperature fluctuation of the accretion disk due to axion heating, as a function of the radius, for the $|211 \rangle$ (blue) and $|322 \rangle$ (red) states. The solid lines present the full numerical results, while the dashed lines present the non-relativistic approximation.  }
    \label{fig:deltat}
\end{figure}

\cref{fig:deltat} shows the temperature fluctuation profile of the disk for the $| 211 \rangle$ and $|322 \rangle$ states. The results for a BH with saturated spin and with the relativistic axion cloud profile (solid), is compared to that of using non-relativistic profile and setting the relativistic factors to unity (dashed). For $\alpha \ll 1$, the cloud is localized far away from the BH horizon, so the effects of the BH spin is negligible. The radius of $(\delta T_c/T_c)_{\rm max}$ is found in this approximation to be $\tilde{r}_{\rm max} \simeq n^{2} \tilde{r}_{c}$, with $(\delta T_c/T_c)_{\rm max}=0.1,\,0.05$ for the $|211 \rangle$ and $| 322 \rangle$ states for benchmark parameters  
(See Supplemental Material for details). 
Notably, as $\alpha$ increases, the temperature fluctuation peaks at smaller radii, as the size of the cloud shrinks.

\section{Effects on X-ray Continuum Spectrum}
\label{sec: effects on x-ray spectrum}
In the thin disk model, thermal photons radiated from the disk follow blackbody radiation $B_{\omega_{e}}(T)\equiv(4\pi^3)^{-1}\omega_{e}^3/(e^{\omega_{e}/T}-1)$ locally. 
According to the MCD model, the flux observed by an observer from a distance $D$ to the XRB with a line-of-sight inclination angle $i$ is 
\begin{equation}
\label{eq:Fd1}
    F_o(\omega)= \frac{1}{2 \pi f^{4}_{\rm col}} \int g^{3} B_{\omega_e}\left(f_{\rm col}(T_{\rm eff}+\delta T_{\rm eff})\right) d \Omega_{\rm obs}\,,
\end{equation}
where the emitted photon frequency is related to the observed frequency via
$\omega_{\rm e}\equiv\omega/g$, 
where $g=g(r_{e},\varphi)$ is the gravitational redshift factor of a photon emitted from a radius $r_{e}$ and an azimuthal angle $\varphi$ in the accretion disk \cite{Cunningham:1975zz}, $f_{\rm col} \simeq1.7$ is the spectral hardening factor that characterizes the heating of thermal photons by Compton scattering with plasma electrons around the disk \cite{Merloni:1999pe}, which is assumed to be constant throughout the disk, $d \Omega_{\rm obs}$ is the solid angle element subtended by the observer,
and the pure-disk surface temperature profile measured by a locally corotating observer is
\begin{equation}
    T_{\rm eff}(r)=\left(\frac{F_{\rm out}(r)}{\sigma_{\rm SB}}\right)^{1/4},
    \label{eq: Teff expr}
\end{equation}
where $F_{\rm out}(r)=\frac{4}{3\tau}\sigma_{\rm SB}T_c^4$ is the outgoing energy flux of the disk without axion heating.
$T_{\rm eff}$ is maximized at $T_{\rm peak}=T_{\rm eff}(r=r_{\rm peak})$.
In the NT disk model, $r_{\rm peak}$ and $T_{\rm peak}$ are determined by \cref{eqn:flux,eq: Teff expr} numerically.
The fluctuation of the surface temperature due to axion heating is obtained through radiative transfer 
\begin{equation}
    \delta T_{\rm eff}=\left(\frac{4}{3\tau}\right)^{1/4} \delta T_c.
\end{equation}
We provide details on how the integral in \Eq{eq:Fd1} is calculated in the Supplemental Material.

\cref{fig:mcd_spec} (left) shows the thermal spectrum of an accretion disk with our benchmark parameters 
as a function of $x_{\rm peak}\equiv\omega/T_{\rm peak}$. For the benchmark model parameters we chose, the peak temperature $ T_{\rm peak}  \simeq 0.25 \textrm{keV}$.
The relative ratio of the peak amplitude between these two spectra is 14\%.
\cref{fig:mcd_spec} (right) shows the shift of the peak frequency $\delta\omega$ with respect to the pure-disk spectrum peak frequency $\omega_*$. The solid lines present the frequency shift of the total spectrum $F_o$ ($\delta\omega_{\rm tot}$, see the calculation in supplemental material), while the dashed lines present that of the spectrum fluctuation $\delta F_o\equiv F_o-F_o(\delta T_{\rm eff}=0)$ only ($\delta\omega_{\rm AH}$). $\delta\omega_{\rm AH}$ shows a monochromatic increase as $\alpha$ increases because the shrinked cloud will heat the higher temperature region.
However, $\delta\omega_{\rm tot}$, the real observable, depends on the amplitude of $\delta F_o$, and thus on $g_{ae}$, and approaches zero at both ends as $q_c\rightarrow 0$.
A non-trivial zero-point location of $\delta\omega$ at $\alpha_0\simeq0.2\,(0.28)$ for $|211\rangle\,(|322\rangle)$ state corresponds to the Bohr radius at $\tilde{r}\simeq 100 $, the radius where the local pure-disk spectrum offers the same peak location as the MCD spectrum. As a result, the spectrum peak frequency experiences a redshift or blueshift for given $\alpha$. Besides, the sudden raise shown in $\delta\omega_{\rm tot}$ corresponds to the additional contribution from the inner region.

\begin{figure*}[h!t]
    \centering
    \includegraphics[width=0.48\linewidth]{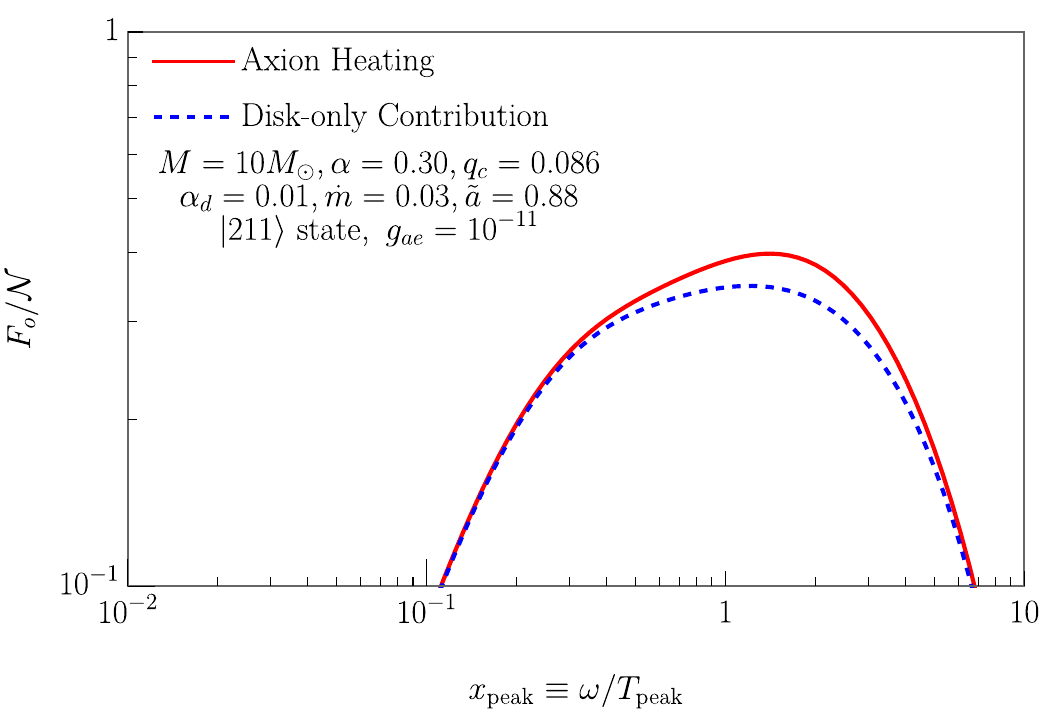}\quad \includegraphics[width=0.48\linewidth]{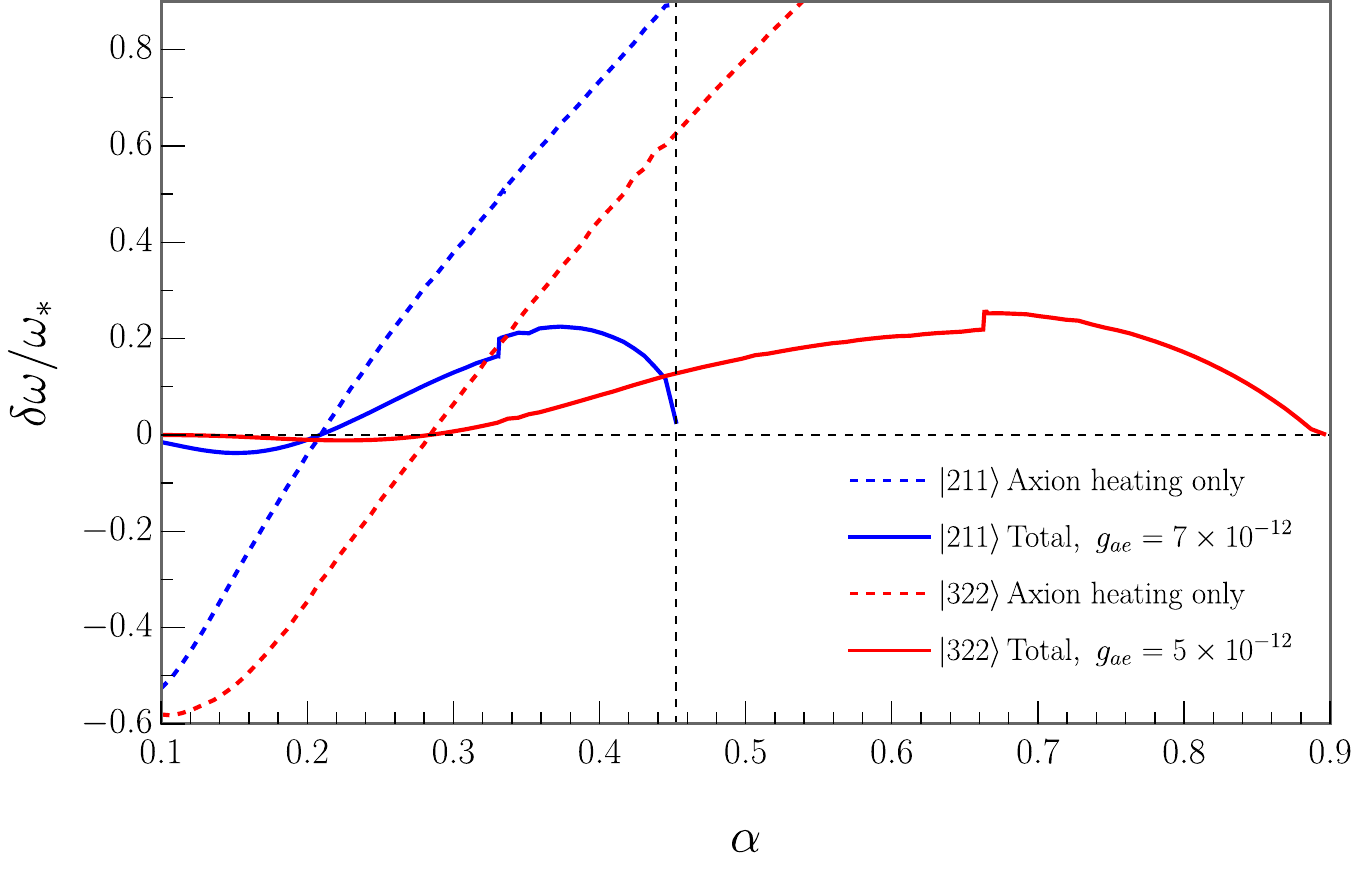}
    \caption{(Left) The normalized observed flux of the MCD thermal spectrum with (red) and without (blue) axion heating, with $\mathcal{N}= r_{\rm in}^2T_{\rm peak}^3/(6 f_{\rm col}D^2)$ and $r_{\rm in}$ the innermost radius of the disk.
    The benchmark model parameters are shown in the figure, and the inclination angle is $i=60^{\degree}$. (Right) Relative frequency shift of the flux peak: axion heating only (dashed), and total (solid).}
    \label{fig:mcd_spec}
\end{figure*}

\section{Discussion}

\cref{fig:MCD_spec_bm1} (right) shows the projected constraints on the axion-electron coupling $g_{ae}$ for both $| 2 1 1 \rangle$ and $| 3 2 2 \rangle$ states and two BH masses $M=3.3,\,10\, M_{\odot}$\footnote{One can recast the $g_{ae}$ bound to the $g_{ap}$ bound by a simple kinematic conversion.}, which is derived by requiring the maximum ratio of $F_o$ between axion heating and disk-only contribution not exceed the uncertainty of the spectral measurement by 10\% or 1\%.  
The initial BH spin is chosen as $\tilde{a}_{\rm ini} = 0.995$, while the spin observed today, $\tilde{a}_{\rm f}$, is set to the superradiant threshold for each $m_{a}$. 
The $m_a$ regime is set by the range of $\alpha$, in which $\alpha_{\rm min}=0.1$ to make the growth timescale of the axion cloud within the typical age of BHs, and $\alpha_{\rm max}=0.445\,(0.896)$ for $|211\rangle(|322\rangle)$ state such that the superradiance condition is satisfied. 
Depending on the uncertainty and the BH mass, our constraints can be tighter than those from the XENONnT experiment\,\cite{XENON:2022ltv}, while they are complementary to those from solar neutrinos\,\cite{Gondolo:2008dd}. A more careful analysis of the continuum fitting can further improve the sensitivity, which we defer for future study.

Below we comment on the axion cloud stability under the axion self-coupling.
Assuming $g_{ae} = C m_{e}/f_{a}$ \cite{DiLuzio:2020wdo},  where $C$ is an $\mathcal{O} \left( 1 \right)$ number and $f_{a}$ is the symmetry breaking scale, the constraints can reach $f_{a} \sim 10^9 \rm GeV$. If the axion self-interaction, arising from a cosine potential, is considered, the cloud cannot grow for such a value of $f_{a}$ \cite{Baryakhtar:2020gao,Arvanitaki:2014wva}. This issue can be evaded if models that suppress self-interactions, such as those with a confinement tower \cite{Agrawal:2017cmd} or axion monodromy \cite{Hebecker:2015zss}, are considered.

The BH spin measurement of XRB is related to the measurement of the innermost stable orbit $r_{\rm in}$, which can be inferred from the continuum fitting of the thermal spectrum \cite{Reynolds:2020jwt}. The smaller $r_{\rm in}$ is, the hotter $T_{\rm peak}$ or the larger peak photon energy $\omega_*$ is. \cref{fig:mcd_spec} demonstrates that the axion heating effect can blueshift or redshift $\omega_*$ by $\delta\omega$, such that the inferred BH spin from the thermal spectrum can be affected. As a result, the axion heating effect may affect some of the BH spin-down constraints on $m_a-f_a^{-1}$, which rely on the BH spin measurement of XRB.
The search of axion heating signal based on the existing observations of BH XRBs is thus worth of detailed studies.

The axion heating effect around a supermassive BH (SMBH) can be studied similarly. However, the Bohr radius of the axion cloud is generically misaligned with the accretion disk of SMBH, which we expect the heating effect to be suppressed compared to the BH XRB.

\section{Acknowledgments} The authors thank Fangzheng Shi for assistance in running \texttt{XPEC}, Rodolfo Capdevilla and Hongwan Liu for useful discussions at the early stage of this project.  The research of F.Y. is supported by the NSF Grant Number PHY-2412701.

\bibliography{ref}

\clearpage

\onecolumngrid

\newpage


\widetext
 \begin{center}
   \textbf{\large SUPPLEMENTAL MATERIAL \\[.2cm] ``Heating Up the Black Hole X-ray Binary Accretion Disk by Superradiance''}\\[.2cm]
  \vspace{0.05in}
  {Antonios Kyriazis, Fengwei Yang, and Siyu Zhou}
\end{center}
\setcounter{equation}{0}
\setcounter{figure}{0}
\setcounter{table}{0}
\setcounter{page}{1}
\setcounter{section}{0} 
\makeatletter
\renewcommand{\thesection}{S-\Roman{section}}
\renewcommand{\theequation}{S-\arabic{equation}}
\renewcommand{\thefigure}{S-\arabic{figure}}

\section{Derivation of the scattering amplitude}

In this section, we present the derivation of the scattering amplitude of the inverse bremsstrahlung process. We can straightforwardly write down the following scattering amplitude for $a+{\rm e}+{\rm p}\rightarrow{\rm e}+\rm{p}$:
\begin{align}
    i\mathcal{M}_{\rm ep}=&\bar{u}\left(p_3\right)\left(-ie\gamma^\mu\right)\frac{i\left(\slashed{p}_1+\slashed{k}+m_{\rm e}\right)}{\left(p_1+k\right)^2-m_{\rm e}^2}\frac{g_{a\rm e}}{2m_{\rm e}}\slashed{k}\gamma_5u\left(p_1\right)D_{\mu\nu}\left(p_4-p_2\right)\bar{u}\left(p_4\right)\left(-ie\gamma^\nu\right)u\left(p_2\right)\nonumber\\
    &+\bar{u}\left(p_3\right)\frac{g_{a\rm e}}{2m_{\rm e}}\slashed{k}\gamma_5\frac{i\left(\slashed{p}_3-\slashed{k}+m_{\rm e}\right)}{\left(p_3-k\right)^2-m_{\rm e}^2}\left(-ie\gamma^\mu\right)u\left(p_1\right)D_{\mu\nu}\left(p_4-p_2\right)\bar{u}\left(p_4\right)\left(-ie\gamma^\nu\right)u\left(p_2\right),
\end{align}
where $D_{\mu\nu}$ is the photon propagator. As we mentioned in the main text, the only effect that the plasma effective mass $\omega_{\rm eff}$ has, is to provide the infrared cut-off for the process. The final result will depend only logarithmically on $\omega_{\rm eff}$. We can also omit the axion energy and apply the non-relativistic limit on fermions. With the above simplifications, we can obtain the squared amplitude:
\begin{align}
\label{appeqn:amplitude}
\begin{split}
    \frac{1}{4}\sum_{\rm spins}\lvert\mathcal{M}_{\rm ep}\rvert^2 
    & \simeq \frac{4 e^4 {g}_{a\rm e}^2 m_{\rm p}^2 \left( \beta_{f} - \beta_{i}\right)^{2}}{ \left( |\textbf{q}|^2 +\omega_{\rm eff} ^{2} \right)^2} \\ & 
    \simeq \frac{4 e^{4} {g}^{2}_{ae} m^{2}_{p}|\textbf{q}|^{2}}{m^{2}_{e} \left( |\textbf{q}|^{2} + \omega_{\rm eff} ^{2} \right)^2}\,,
    \end{split}
\end{align}
where $\beta_{i}$ and $\beta_{f}$ are the initial and final velocities of the electrons, respectively.  We may therefore write $m^{2}_{e} (\beta_{f} - \beta_{i})^{2} = (p_{f} - p_{i})^{2} \approx |\textbf{q}|^{2}$. In deriving \cref{appeqn:amplitude}, we consider that the typical momentum of the electrons in the plasma with a temperature $T$ is $|\textbf{p}_{e,i(f)}| \sim \sqrt{m_{e} T} \gg |\textbf{k}_{a}|$, which is valid in the disk, where typical temperatures are in the range $T \sim 10^{6} - 10^{7} \,$K, while the momentum of the axion is of order $|\textbf{k}_{a}| \sim m_{a} \alpha$. 

To motivate the form of the amplitude in \cref{appeqn:amplitude}, let us recall that in the non-relativistic approximation, a Dirac spinor scales as $\psi \sim \sqrt{m_\psi}$, where $m_\psi$ is the mass of the fermion \cite{Peskin:1995ev}. This explains the appearance of $ m^{2}_{p}$. $\omega_{\rm eff}$ reflects the modified dispersion relation of photons in the plasma.  
Finally, $(\beta_{f} - \beta_{i})^{2}$ can be anticipated by the fact that bremsstrahlung occurs only if the electron's velocity changes. Indeed, Larmor's formula for emission of light by a charged particle in motion depends on the particle's acceleration \cite{Jackson:1998nia}; therefore, its velocity must be changing if it is to radiate.   

Applying a similar approach on $a+{\rm e}+{\rm e}\rightarrow{\rm e}+\rm{e}$ process, we get
\begin{equation}
    \frac{1}{2}\frac{1}{4}\sum_{\rm spins}\lvert\mathcal{M}_{\rm ee}\rvert^2\simeq \frac{4 e^{4} {g}^{2}_{ae}|\textbf{q}|^{2}}{\left( |\textbf{q}|^{2} + \omega_{\rm eff} ^{2} \right)^2}\,,
\end{equation}
The extra $1/2$ factor comes from the average over identical electrons in final state. 

\section{Emissivity integration}
In this section, we present a detailed derivation of the emissivity, starting from the general formula:
\begin{align}
    \epsilon_a&=\int {\rm d}\Pi_aE_a{\rm d}\Pi_1{\rm d}\Pi_2{\rm d}\Pi_3{\rm d}\Pi_4 (2\pi)^4\delta^{(4)}(p_1+p_2+k-p_3-p_4)|\mathcal{M}|^2 f_af_1f_2(1-f_3)(1-f_4)\nonumber\\
    &\quad-\int {\rm d}\Pi_aE_a{\rm d}\Pi_1{\rm d}\Pi_2{\rm d}\Pi_3{\rm d}\Pi_4 (2\pi)^4\delta^{(4)}(p_3+p_4+k-p_1-p_2)|\mathcal{M}|^2 f_3f_4(1-f_1)(1-f_2)(1+f_a)\nonumber\\
    &\approx\frac{1}{T}\int {\rm d}\Pi_aE_a^2{\rm d}\Pi_1{\rm d}\Pi_2{\rm d}\Pi_3{\rm d}\Pi_4 (2\pi)^4\delta^{(4)}(p_1+p_2+k-p_3-p_4)|\mathcal{M}|^2 f_af_1f_2(1-f_3)(1-f_4)
    \label{appeq: general emmisivity}
\end{align}
where the first line is the inverse bremsstrahlung that heats up the plasma, and the second line is its inverse process that cools down the plasma. It is evident that $f_a\gg1$, so we can take $1+f_a\rightarrow f_a$. 

As far as the fermions are concerned, we shall assume a Maxwell-Boltzmann distribution:
\begin{equation}
f_i(\vecp)=n_i\left(\frac{2\pi}{m_iT}\right)^{\frac{3}{2}}e^{-\frac{p^2}{2m_iT}}\,,
\end{equation}
where $n_i$ is the number density. In this approximation, it holds that $f_i\ll1$ and therefore $1-f_i\approx1$.
One can easily check that the cooling process is $e^{-E_a/T}$ times of the heating process by applying $\vecp_i\rightarrow-\vecp_i$ and then the energy conservation condition. As $E_a\ll T$, the exponential factor can be Taylor expanded and we obtain the third line of Eq.~(\ref{appeq: general emmisivity}).

In the disk, fermions are non-degenerate, i.e., $f_i, f_f \ll1$, so it is a good approximation to use the Boltzmann distribution for the electrons and ignore Pauli blocking. The number density of the superradiant cloud is so large, $\bar{n}_{a} 
    \sim 3 \times 10^{50} \textrm{cm}^{-3} \left(\alpha/0.1\right)^{5} \left( M_{\odot}/M\right)$, that one can approximate in Eq.~(\ref{eqn: emissivity definition}) $f_{a} \gg 1$ and assume a form $f_{a} = (2 \pi)^{3} n_{a} \delta^{(3)}(\textbf{k}_{a})$, since the axion momentum can be safely ignored in the calculations when compared to the other parameters.

In addition, since axion energy is negligible, we can omit the $k^\mu$ dependence in $\delta$-function. Thus, the axion momentum integration can be decoupled from the fermions' momentum integration. Now we can simplify the emissivity as
\begin{equation}
    \epsilon_a\approx\frac{\rho_a}{2T}\int{\rm d}\Pi_1{\rm d}\Pi_2{\rm d}\Pi_3{\rm d}\Pi_4 (2\pi)^4\delta^{(4)}(p_1+p_2-p_3-p_4)|\mathcal{M}|^2 f_1f_2\,,
    \label{eq: emissivity, axion decoupled}
\end{equation}
where $\rho_a$ is the axion energy density, which can be approximate as $n_am_a$.

As we've shown in the above section, the squared amplitude for the $a+{\rm e}+{\rm p}\rightarrow{\rm e}+\rm{p}$ and $a+{\rm e}+{\rm e}\rightarrow{\rm e}+\rm{e}$ processes can be written in the form
\begin{equation}
    \frac{4 e^{4} {g}^{2}_{ae} m^{2}_{2}|\textbf{q}|^{2}}{m^{2}_{\rm e} \left( |\textbf{q}|^{2} + \omega_{\rm eff} ^{2} \right)^2}\,.
\end{equation}
Hence, here we do the calculation for both cases together by using $i$ indices, instead of e and p. Also, note that $m_1=m_3$, $m_2=m_4$. As in the kinematics of 2-body elastic scattering in the center of mass frame, we define $M_{12}=m_1+m_2$, the reduced mass $\mu=m_1m_2/(m_1+m_2)$, the total momentum $\boldsymbol{P}=\vecp_1+\vecp_2$ and the relative momentum $\vecp=(m_2\vecp_1-m_1\vecp_2)/M_{12}$. We apply the same notation for final states with $\boldsymbol{P}^\prime$ and $\vecp^\prime$. For elastic scattering, we have $\lvert\vecq\rvert^2=2p^2(1-\cos\theta)$, where we've used the momentum conservation and $\theta$ is the angle between $\vecp$ and $\vecp^\prime$. Now, we are ready to calculate Eq.~(\ref{eq: emissivity, axion decoupled})
\begin{align}
    \epsilon_a&\approx\frac{\rho_a}{2T}\int{\rm d}\Pi_P{\rm d}\Pi_p{\rm d}\Pi_{P^\prime}{\rm d}\Pi_{p^\prime} (2\pi)^3\delta^{(3)}(\boldsymbol{P}-\boldsymbol{P}^\prime)(2\pi)\delta\left(\frac{P^2}{2M_{12}}+\frac{p^2}{2\mu}-\frac{P^{\prime2}}{2M_{12}}-\frac{p^{\prime2}}{2\mu}\right)|\mathcal{M}|^2 f_Pf_p\nonumber\\
    &=\frac{2\rho_am_2^2}{Tm_{\rm e}^2}e^4g_{a\rm e}^2n_{1}n_{2}\left(\frac{2\pi}{\mu T}\right)^{\frac{3}{2}}\frac{1}{(2M_{12})^2(2\mu)^2}\int\frac{{\rm d}^3p}{(2\pi)^3}\frac{{\rm d}^3p^\prime}{(2\pi)^3}(2\pi)\delta\left(\frac{p^2}{2\mu}-\frac{p^{\prime2}}{2\mu}\right)\frac{2p^2(1-\cos\theta)}{\left[ 2p^2(1-\cos\theta)+\omega_{\rm eff}^2\right]^2}e^{-\frac{p^2}{2\mu T}}
     \nonumber\\
     &=\frac{\rho_am_2^2}{32\pi^3M_{12}^2\mu Tm_{\rm e}^2}e^4g_{a\rm e}^2n_{1}n_{2}\left(\frac{2\pi}{\mu T}\right)^{\frac{3}{2}}\int{\rm d}p{\rm d}\cos\theta\frac{2p^5(1-\cos\theta)}{\left[ 2p^2(1-\cos\theta)+\omega_{\rm eff}^2\right]^2}e^{-\frac{p^2}{2\mu T}}\nonumber\\
     &=\frac{\rho_am_2^2}{32\pi^3M_{12}^2m_{\rm e}^2}e^4g_{a\rm e}^2n_{1}n_{2}\left(\frac{2\pi}{\mu T}\right)^{\frac{3}{2}}\frac{\mu}{T}\int{\rm d}v{\rm d}\cos\theta\frac{2v^5(1-\cos\theta)}{ \left[2v^2(1-\cos\theta)+\omega_{\rm eff}^2/\mu^2\right]^2}e^{-\frac{\mu v^2}{2T}}\nonumber\\
     &=\frac{\rho_am_2^2}{32\pi^3M_{12}^2m_{\rm e}^2}e^4g_{a\rm e}^2n_{1}n_{2}\left(\frac{2\pi}{\mu T}\right)^{\frac{3}{2}}\left(-\frac{1}{2}\right)\left[1+e^x{\rm Ei}(-\zeta)-\zeta\Gamma(0,\zeta)\right],\qquad \zeta\equiv\frac{\omega_{\rm eff}^2}{8T\mu}\ll1\nonumber\\
     &\equiv\frac{\rho_am_2^2}{32\pi^3M_{12}^2m_{\rm e}^2}e^4g_{a\rm e}^2n_{1}n_{2}\left(\frac{2\pi}{\mu T}\right)^{\frac{3}{2}}F(\zeta)\,,
\end{align}
where $f_Pf_p$ in the first line is the distribution function with number density unchanged and other quantities accordingly changed to center of mass frame from $f_1f_2$. In the fifth line, ${\rm Ei}(\zeta)\equiv-\int_{-\zeta}^\infty e^{-t}/t{\rm d}t$ is the exponential integral function and $\Gamma(\vartheta,\zeta)\equiv\int^{\infty}_{\zeta}t^{\vartheta-1}e^{-t}{\rm d}t$ is the incomplete Gamma function. In the last line, we defined the form factor, which in the $\zeta\rightarrow0$ limit, is approximated as
\begin{equation}
    F(\zeta)=-\frac{1}{2}\left(1+\gamma+\ln{\zeta}\right),
\end{equation}
where $\gamma$ is the Euler constant. 

Now, we add the contribution from both $a+{\rm e}+{\rm p}\rightarrow{\rm e}+\rm{p}$ and $a+{\rm e}+{\rm e}\rightarrow{\rm e}+\rm{e}$ processes. Note that $F(\zeta)$ us weakly related to $\mu$. Hence, $a+{\rm e}+{\rm p}\rightarrow{\rm e}+\rm{p}$ contributes roughly  $\sqrt{2}$ times of $a+{\rm e}+{\rm e}\rightarrow{\rm e}+\rm{e}$ process. Neglecting the small difference of $F(\zeta)$, we get the total heating rate per baryon
\begin{equation}
    \frac{\epsilon_a}{n_{\rm p}}=0.85
    \left(\frac{g_{a\rm e}}{10^{-11}}\right)^2\frac{E_an_a}{10^{41}{\rm eV}{\rm cm}^{-3}}\frac{n_{\rm e}}{10^{22}{\rm cm}^{-3}}\left(\frac{10^7 {\rm K}}{T}\right)^{\frac{3}{2}}F(\zeta)\,{\rm eV/s}\,,
\end{equation}
where we have assumed that the plasma is formed purely by protons and electrons, such that $n_{\rm e}=n_{\rm p}$.

\section{Summary of the Novikov-Thorne disk model}

In the {\it inner region}:
\begin{align}
\label{appeq:inner_region_profile}
   \rho_{\rm pl} &\simeq 2\times 10^{-5}\text{g/cm}^3\alpha_d^{-1}\tilde{m}^{-1}\dot{m}^{-2} \tilde{r}^{3/2} \frac{\mathcal{B}^{6}\mathcal{D}\mathcal{E}^2}{\mathcal{A}^{4}\mathcal{Q}^{2}}\, ,\nonumber\\
   T_{\rm c}&\simeq5\times 10^7\,{\rm K}\,\alpha_d^{-\frac{1}{4}}\tilde{m}^{-1/4}\tilde{r}^{-3/8}\frac{\mathcal{B}^{1/2}\mathcal{E}^{1/4}}{\mathcal{A}^{1/2}},\nonumber\\
   \Sigma&\simeq5\text{g/cm}^2\,\alpha_d^{-1}\dot{m}^{-1}\tilde{r}^{3/2}\frac{\mathcal{B}^3\mathcal{C}^{1/2}\mathcal{E}}{\mathcal{A}^2\mathcal{Q}},
\end{align}
and in the {\it middle region}:
\begin{align}
\label{appeq:middle_region_profile}
   \rho_{\rm pl} &\simeq 40\,\text{g/cm}^3\alpha_d^{-\frac{7}{10}}\tilde{m}^{-\frac{7}{10}}\dot{m}^{2/5} \tilde{r}^{-\frac{33}{20}} \frac{\mathcal{B}^{3/5}\mathcal{E}^{1/2}\mathcal{Q}^{2/5}}{\mathcal{A}\mathcal{D}^{1/5} },\nonumber\\
   T_{\rm c}&\simeq7\times 10^8\,{\rm K}\,\alpha_d^{-\frac{1}{5}}\tilde{m}^{-\frac{1}{5}}\dot{m}^{2/5}\tilde{r}^{-\frac{9}{10}}\frac{\mathcal{Q}^{2/5}}{\mathcal{B}^{2/5}\mathcal{D}^{1/5}},\nonumber\\
   \Sigma&\simeq9\times10^4\text{g/cm}^2\,\alpha_d^{-4/5}\tilde{m}^{1/5}\dot{m}^{3/5}\tilde{r}^{-3/5}\frac{\mathcal{C}^{1/2}\mathcal{Q}^{3/5}}{\mathcal{B}^{4/5}\mathcal{D}^{4/5}},
\end{align}
where $\tilde{m}\equiv M/M_\odot$, $\tilde{r}\equiv r/r_g$,
$\mathcal{A}=1+\tilde{a}^2x^{-4}+2\tilde{a}^2x^{-6},\,\mathcal{B}=1+\tilde{a}x^{-3},\,\mathcal{C}=1-3x^{-2}+2\tilde{a}x^{-3},\,\mathcal{D}=1-2x^{-2}+\tilde{a}^2x^{-4},\,\mathcal{E}=1+4\tilde{a}^2x^{-4}-4\tilde{a}^2x^{-6}+3\tilde{a}^4x^{-8},\,\,x\equiv\sqrt{r/r_g}$, are Kerr metric coefficients, $\tilde{a}=a/M$ is the dimensionless BH spin, $\mathcal{Q}$ is the relativistic Page-Thorne flux factor encoding the boundary condition for the disk profile \cite{Page:1974he},
\begin{eqnarray} 
   \mathcal{Q}(x)&=&\mathcal{Q}_0\bigg[x-x_0-\frac{3}{2}\tilde{a}\ln\left(\frac{x}{x_0}\right)-\frac{3(x_1-\tilde{a}^2)}{x_1 (x_1-x_2)(x_1-x_3)}\ln\left(\frac{x-x_1}{x_0-x_1}\right)\nonumber\\
   &-&\frac{3(x_2-\tilde{a}^2)}{x_2 (x_2-x_3)(x_2-x_1)}\ln\left(\frac{x-x_2}{x_0-x_2}\right)-\frac{3(x_3-\tilde{a}^2)}{x_3 (x_3-x_1)(x_3-x_2)}\ln\left(\frac{x-x_3}{x_0-x_3}\right)\bigg], 
\end{eqnarray}
where 
\begin{eqnarray}
    &\mathcal{Q}_0=\frac{1+\tilde{a}x^{-3}}{x(1-3x^{-2}+2\tilde{a}x^{-3})^{1/2}},\nonumber\\
   & x_0=\sqrt{r_{\rm in}/r_g}, ~~x_1=2\cos[(\cos^{-1}\tilde{a}-\pi)/3],\nonumber\\
&x_2=2\cos[(\cos^{-1}\tilde{a}+\pi)/3],~x_3=-2\cos[(\cos^{-1}\tilde{a})/3],
\end{eqnarray} 
with $r_{\rm in}$ the radius of the innermost stable circular geodesic orbit. 

By comparing \Cref{appeq:inner_region_profile,appeq:middle_region_profile} with \Cref{eq:inner_rho,eq:inner_Tc,eq:inner_Sigma,eq:middle_rho,eq:middle_Tc,eq:middle_Sigma}, the definition of auxiliary functions is
\begin{eqnarray}
    h_1=\frac{\mathcal{B}^{6}\mathcal{D}\mathcal{E}^2}{\mathcal{A}^{4}\mathcal{Q}^{2}},h_2=\frac{\mathcal{B}^{1/2}\mathcal{E}^{1/4}}{\mathcal{A}^{1/2}},h_3=\frac{\mathcal{B}^3\mathcal{C}^{1/2}\mathcal{E}}{\mathcal{A}^2\mathcal{Q}},
    \nonumber\\
    h_4=\frac{\mathcal{B}^{3/5}\mathcal{E}^{1/2}\mathcal{Q}^{2/5}}{\mathcal{A}\mathcal{D}^{1/5} },h_5=\frac{\mathcal{Q}^{2/5}}{\mathcal{B}^{2/5}\mathcal{D}^{1/5}},h_6=\frac{\mathcal{C}^{1/2}\mathcal{Q}^{3/5}}{\mathcal{B}^{4/5}\mathcal{D}^{4/5}}.
\end{eqnarray}

The peak radius $r_{\rm peak}$ is defined as the position where the effective surface temperature $T_{\rm eff}(r)$ is maximized. So, given the profile of $T_{\rm eff}(r)$, one can determine $r_{\rm peak}$ by calculating its extrema:
\begin{equation}
    \frac{\partial T_{\rm eff}}{\partial r}\bigg\vert_{r=r_{\rm peak}}=0,
\end{equation}
where the surface temperature profile can be obtained by combining \cref{eqn:flux} and 
\begin{equation}
    T_{\rm eff}=\left(\frac{4}{3\tau}\right)^{1/4}  T_{\rm c},
\end{equation}
\begin{equation}
    \Rightarrow T_{\rm eff}(r)\propto r^{-3/4} \mathcal{Q}^{'1/4}(r).
\end{equation}

Combining Eq.\,\eqref{eq: energy per electron} with the disk model profiles, e.g., \cref{eq:middle_rho,eq:middle_Tc,eq:middle_Sigma} ({\it middle region}), and consider $\alpha\lesssim0.3$, we can determine the axion emissivity as
\begin{align}
\label{eq:emiss radial}
    \left( \frac{\epsilon_{a}}{n_{e}} \right)_{nlm} & = \left( \frac{\epsilon_{a}}{n_{e}} \right)^{\ast}_{nlm} \mathcal{Z}_{nlm}(\tilde{r}) F(\zeta)  h_{4}(x,\tilde{a}) h^{-3/2}_{5} (x,\tilde{a})\,,
\end{align}
with 
\begin{align}
\begin{split}
    \left( \frac{\epsilon_{a}}{n_{e}} \right)^{\ast}_{211} \simeq 92 \frac{\textrm{eV}}{s} \left( \frac{\dot{m}}{0.03} \right)^{-1/5} & \left( \frac{M}{10 M_{\odot}} \right)^{-12/5} \left( \frac{\alpha_{d}}{0.01} \right)^{-2/5} \left( \frac{g_{ae}}{ 10^{-11}} \right)^{2} \left( \frac{\alpha}{0.3} \right)^{33/5} \left( \frac{q_{c,211}}{0.086} \right), \\  
     \left( \frac{\epsilon_{a}}{n_{e}} \right)^{\ast}_{322} & \simeq 2.43 \times 10^{-3}  \left( \frac{q_{c,322}}{q_{c,211}} \right) \left( \frac{\epsilon_{a}}{n_{e}} \right)^{\ast}_{211}\,,
\end{split}
\end{align}
and 
\begin{align}
    \begin{split}
        \mathcal{Z}_{211}(\tilde{r},\tilde{r}_{c}) = 64 \pi \tilde{r}^{3}_{c} \left( \frac{\tilde{r}}{\tilde{r}_{c}} \right)^{-3/10} |\tilde{\psi}_{211}|^{2}\,\qquad
        \mathcal{Z}_{322}(\tilde{r},\tilde{r}_{c}) = 162^{2} \pi \tilde{r}^{3}_{c} \left( \frac{\tilde{r}}{\tilde{r}_{c}} \right)^{-3/10} |\tilde{\psi}_{322}|^{2}\,.
    \end{split}
\end{align}
In the non-relativistic limit, using the wavefunctions \cref{eq:wavefunctions NR}, we obtain for the $\mathcal{Z}_{nlm}$ functions:
\begin{align}
\begin{split}
    \mathcal{Z}_{211}(\tilde{r},\tilde{r}_{c}) = e^{-\tilde{r}/\tilde{r_{c}}}  \left( \frac{\tilde{r}}{\tilde{r}_{c}} \right)^{17/10},\qquad 
     \mathcal{Z}_{322}(\tilde{r},\tilde{r}_{c}) = e^{-2 \tilde{r}/ 3 \tilde{r}_{c}}  \left( \frac{\tilde{r}}{\tilde{r}_{c}} \right)^{37/10}.
\end{split}
\end{align}
Note that $\zeta$ also depends on $\tilde{r}$ through $T_{c}$ and $\rho_{\rm pl}$, albeit very mildly.

\subsection{Temperature fluctuation}

From \Eq{eqn:temp fluct} and \Eq{appeq:inner_region_profile}, \Eq{appeq:middle_region_profile}, we can straightforwardly obtain the radial profile of the temperature fluctuation,
\begin{align}
\begin{split}
   \left( \frac{\delta T_{c}}{T_{c}} \right)_{nlm} = \left( \frac{\delta T_{c}}{T_{c}} \right)^{\ast}_{nlm}  \mathcal{W}_{nlm}(\tilde{r}) F(\zeta)  h_{4}  h_{5}^{-5.5}  h_{6}^{2}\,,
   \label{eq: deltaT/T}
\end{split}
\end{align}
where we separate the model-parameter dependence
\begin{align}
    \left( \frac{\delta T_{c}}{T_{c}} \right)^{\ast}_{211} & = 1.62 \times 10^{-3} \left( \frac{\alpha}{0.3} \right)^{\frac{9}{5}} \left( \frac{M}{10 M_{\odot}}\right)^{-\frac{6}{5}} \left( \frac{\dot{m}}{0.03}\right)^{-\frac{3}{5}}  \left( \frac{\alpha_{d}}{0.01} \right)^{-\frac{6}{5}} \left( \frac{g_{ae}}{10^{-11}} \right)^{2} \left( \frac{q_{c,211}}{0.086} \right) , \label{eq: deltaToverT star} \\ 
    \left( \frac{\delta T_{c}}{T_{c}} \right)^{\ast}_{322} & = 2.44 \times 10^{-3}  \left( \frac{\delta T_{c}}{T_{c}} \right)^{\ast}_{211} \left( \frac{q_{c,322}}{q_{c,211}} \right),
\end{align}

and the radial dependence

\begin{align}
    \begin{split}
        \mathcal{W}_{211} (\tilde{r},\tilde{r}_{c}) & = 64 \pi \tilde{r}^{3}_{c} \left( \frac{\tilde{r}}{\tilde{r}_{c}} \right)^{21/10} |\tilde{\psi}_{211}|^{2} \\ 
         \mathcal{W}_{322} (\tilde{r},\tilde{r}_{c}) & = 162^{2} \pi \tilde{r}^{3}_{c} \left( \frac{\tilde{r}}{\tilde{r}_{c}} \right)^{21/10} |\tilde{\psi}_{322}|^{2}
    \end{split}
\end{align}

Using the non-relativistic wavefunctions, we obtain for $\mathcal{W}_{211}$ and $\mathcal{W}_{322}$:

\begin{align}
        \mathcal{W}_{211}  = \left(\frac{\tilde{r}}{\tilde{r}_c}\right)^{4.1} e^{-\frac{\tilde{r}}{\tilde{r}_c}}\,, \quad
         \mathcal{W}_{322}  = \left(\frac{\tilde{r}}{\tilde{r}_c}\right)^{6.1}e^{-\frac{2\tilde{r}}{3\tilde{r}_c}}\,.
\end{align}
For $\alpha\sim 1$, we compute \cref{eq: deltaT/T} by using the relativistic axion cloud profile to derive $\mathcal{W}_{nlm} $.

\subsection{Thermalization time scale}
The axion-heating time scale can be characterized by $\tau_{a} \equiv \Gamma^{-1}_{a}$ with $\Gamma_{a} \equiv \epsilon_{a}/(n_{\rm p} E_{a})$. At the radius where the emissivity is maximized, roughly the Bohr radius of the cloud $\tilde{r} \sim \tilde{r}_c$, $\tau_{a}\simeq 10^{-14}\, \textrm{sec}$ for $M=10M_\odot,\alpha=0.3,\alpha_{d}=0.01,\dot{m}=0.03,g_{ae}=10^{-11}$ for the $|211\rangle$ state.
Compared with the thermal time scale for the heat dissipation across the disk
$t_{\rm th} \sim \alpha^{-1}_{d} r_{g} \tilde{r}^{3/2}$ \cite{Frank_King_Raine_2002}, with the same parameters chosen above, $t_{\rm th} \simeq 0.6 \,\textrm{sec} $.
Since $\tau_{a}\ll t_{\rm th}$, the axion cloud injects heat in the disk sufficiently fast such that the disk can be considered in thermal equilibrium. The same result holds for the $|322 \rangle$ state.

\section{Calculation of the continuous thermal spectrum}

To calculate the observed spectrum from the accretion disk, \cref{eq:Fd1}, we follow the methodology of \cite{Cunningham:1975zz,SPEITH1995109}. This calculation requires first the calculation of the photon geodesics that cross the accretion disk and reach the observer. The photon geodesics are characterized by the dimensionless constants of motion $\lambda$ and $q$, where the former is the angular momentum along the rotation axis of the black hole and the latter is the Carter constant \cite{Carter:1968rr}, both normalized by the energy of the photon. The redshift of the photon is defined by:

\begin{equation}
    \label{eqn:redshift}
    g(r_{e},\lambda,q,i) \equiv \frac{\omega}{\omega_{e}},
\end{equation}
where $\omega$ is the frequency measured by the observer and $\omega_{e}$ is the emitted frequency. We may obtain $\lambda$ in terms of $r_{e}$ and $q$ by using the equations of motion of the photons, where $r_e$ is the photon emitted radius on the disk, as described in \cite{Cunningham:1975zz,SPEITH1995109}. After this reduction, for a given $r_{e}$, the redshift $g$ can be extremized in terms of $q$ to obtain $g_{\rm max}(r_{e},i)$ and $g_{\rm min}(r_{e},i)$.
\begin{figure}
    \centering
    \includegraphics[width=0.45\linewidth]{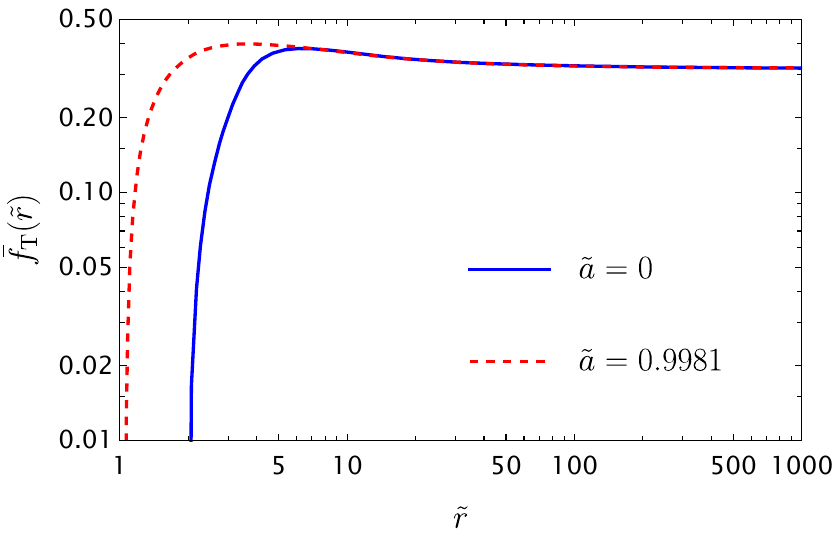}\quad \includegraphics[width=0.45\linewidth]{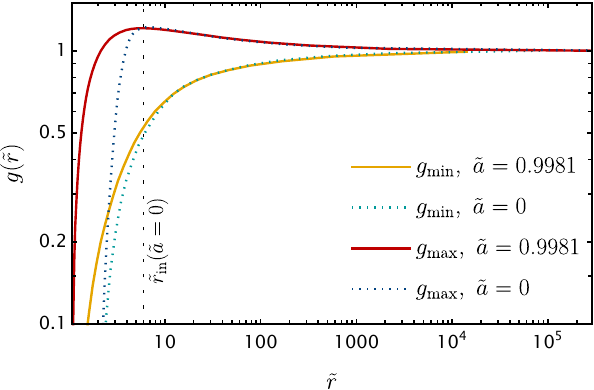}
    \caption{Transfer function $\bar{f}_T$ and redshift factor $g_{\rm max},\,g_{\rm min}$ at $i=60^{ \degree}$.}
    \label{fig:fTandg}
\end{figure}
Turning now to the integral in \cref{eq:Fd1}, the solid angle subtended by the observer is written as:

\begin{equation}
d\Omega_{\rm obs} = \frac{\partial \Omega_{\rm obs}}{\partial \left( g_{\ast},r_{e}\right)} dg_{\ast} dr_{e} = \frac{\pi r_{e}}{D^{2}} \frac{1}{g \sqrt{g_{\ast} \left(1 - g_{\ast} \right)}} f(g_{\ast},r_{e},i) dg_{\ast} dr_{e}
\end{equation}
where the transfer function $f(g_{\ast},r_{e},i)$ is defined as:

\begin{equation}
    f(g_{\ast},r_{e},i) \equiv \frac{D^{2}}{\pi r_{e}} g \sqrt{g_{\ast} \left(1 - g_{\ast} \right)} \frac{\partial \Omega_{\rm obs}}{\partial \left( g_{\ast},r_{e}\right)},
\end{equation}
with $g_{\ast} = \left(g - g_{\rm min} \right)/\left( g_{\rm max} - g_{\rm min} \right)$. We refer again the reader to the appendix of \cite{Cunningham:1975zz} for the calculation of $\partial \Omega_{\rm obs} / \partial \left( g_{\ast},r_{e} \right)$. This transfer function is double valued for $i \neq 90^{o}$, but when the specific intensity is isotropic, as we assume in this work, the transfer function that we use in practice is given by the sum of the two values $f_{T}(g_{\ast},r_{e},i) = f_{1}(g_{\ast},r_{e},i) + f_{2}(g_{\ast},r_{e},i)$. In addition, since the integrand is dominated at $g_{\ast} \approx 1$ and $f_{T}$ is approximately constant as $g_{\ast}$ is varied, we calculate $f_{T}$ at $g_{\ast} =0.95$. The transfer function $f_{T}$ and $g_{\rm max},g_{\rm min}$ are shown as a function of $r_{e}$ in \cref{fig:fTandg} for inclination angle $i=60^{\degree}$ for a maximally spinning and a non-spinning black hole. For this inclination angle, it is clear that changing the spin only affects the ISCO, with the functions remaining self-similar. 
We set the integration range of $r\in[r_{\rm in},r_{\rm BC}]$, but the axion heating effect, i.e., $\delta T_{\rm eff}/T_{\rm eff}\neq0$, is only turned on when $r\ge\max({r_{\rm in},r_{\rm AB}})$, where $r_{\rm AB}$ is the radius that separates the inner region with the middle region and $r_{\rm BC}$ is the radius that separates the middle region with the outer region.

\section{Peak location shift}

In this section, we show how axion heating will influence the peak location of flux from the pure disk. Assume the pure-disk flux peaks at $\omega=\omega_*$,
\begin{equation}
    \left.\frac{\partial F}{\partial\omega}\right|_*=0\,.
    \label{eq: unperturbed peak}
\end{equation}
We here use perturbation again. The heated flux is $F+\delta F$ and new peak is at $\omega_*+\delta\omega$. This gives
\begin{equation}
    \frac{\partial(F+\delta F)}{\partial(\omega_*+\delta\omega)}=\left.\frac{\partial F}{\partial\omega}\right|_*+\left.\frac{\partial^2 F}{\partial\omega^2}\right|_*\delta\omega+\left.\frac{\partial\delta F}{\partial\omega}\right|_*=0\,,
    \label{eq: perturbed peak}
\end{equation}
where we have omit higher-order terms. Substitute Eq.~(\ref{eq: unperturbed peak}) into Eq.~(\ref{eq: perturbed peak}), one can obtain
\begin{equation}
    \delta\omega=-\left.\left.\frac{\partial\delta F}{\partial\omega}\right|_*\right/\left.\frac{\partial^2 F}{\partial\omega^2}\right|_*\,.
    \label{eq: delta omega}
\end{equation}
This is as expected. As shown in \cref{fig:mcd_spec}, both $F$ and $\delta F$ are convex functions, so $\partial^2F/\partial\omega^2|_*<0$, and if the peak location of $\delta F$ is larger than $\omega_*$, $\partial\delta F/\partial\omega$ is larger than 0 at $\omega_*$, which means the direction of peak shifting is consistent with $\delta F$'s peak position.

Now, we can solve the numerator part.
\begin{align}
    \frac{\partial\delta F_o}{\partial\omega}&= \frac{1}{2 \pi f^{4}_{\rm col}}\int g^{3} \frac{\partial^2B_{\omega_e}\left(T_{\rm col}\right)}{\partial\omega\partial T_{\rm col}}\delta T_{\rm col}{\rm d} \Omega_{\rm obs}\,,
    \nonumber\\
    &=\frac{1}{\omega}\int g^3B_{\omega_e}\left(T_{\rm col}\right)\frac{\delta T_{c}}{T_{c}}\left[D_1(x_{\rm e,col})-D_2(x_{\rm e,col})\right]{\rm d} \Omega_{\rm obs}\,,
\end{align}
where in the first line, we defined $T_{\rm col}=f_{\rm col}T_{\rm eff}$, in the second line, we used  $\delta T_{\rm col}/T_{\rm col}=\delta T_{\rm eff}/T_{\rm eff}  = \delta T_{c} / T_{c}$, $x_{\rm e,col}\equiv\omega_e/T_{\rm col}$, and 
\begin{align}
    D_1(x)=\frac{4xe^x}{e^x-1}\,,\qquad
    D_2(x)=\frac{x^2e^x(e^x+1)}{(e^x-1)^2}\,.
\end{align}

\end{document}